\documentclass[aps,prd,twocolumn,noshowpacs,superscriptaddress,groupedaddress,nofootinbib,preprintnumbers]{revtex4}
\pdfoutput=1
\usepackage{amsmath}
\usepackage{amsfonts}
\usepackage{amssymb}
\usepackage{graphicx, rotating}
\usepackage{epstopdf}
\usepackage{epsfig}
\usepackage{empheq}
\usepackage{latexsym}
\usepackage{multirow}
\usepackage{color}
\usepackage{amsmath,amssymb}
\usepackage{slashed}
\usepackage[colorlinks=true,linkcolor=red,anchorcolor=black,citecolor=blue,filecolor=cyan,menucolor=red,runcolor=filecolor,urlcolor=blue,bookmarks=true,bookmarksnumbered=true]{hyperref}%
\definecolor{rossos}{cmyk}{0,1,1,0.55}
\definecolor{bluscuro}{rgb}{0.15, 0.2, .85}
\definecolor{bluchiaro}{cmyk}{1,.3,0.,0.1}

\definecolor{rossos}{cmyk}{0,1,1,0.55}
\definecolor{bluscuro}{rgb}{0.15, 0.2, .85}
\definecolor{bluchiaro}{cmyk}{1,.3,0.,0.1}

\newcommand{\bc}{\begin{center}}
\newcommand{\ec}{\end{center}}
\newcommand{\bea}{\begin{eqnarray}}
\newcommand{\eea}{\end{eqnarray}}

\catcode`@=11
\def\marginnote#1{}

\newcount\hour
\newcount\minute
\newtoks\amorpm
\hour=\time\divide\hour by60
\minute=\time{\multiply\hour by60 \global\advance\minute by-\hour}
\edef\standardtime{{\ifnum\hour<12 \global\amorpm={am}%
\else\global\amorpm={pm}\advance\hour by-12 \fi
\ifnum\hour=0 \hour=12 \fi
\number\hour:\ifnum\minute<10 0\fi\number\minute\the\amorpm}}
\edef\militarytime{\number\hour:\ifnum\minute<10 0\fi\number\minute}
\def\draftlabel#1{{\@bsphack\if@filesw {\let\thepage\relax
\xdef\@gtempa{\write\@auxout{\string
\newlabel{#1}{{\@currentlabel}{\thepage}}}}}\@gtempa
\if@nobreak \ifvmode\nobreak\fi\fi\fi\@esphack}
\gdef\@eqnlabel{#1}}
\def\@eqnlabel{}
\def\@vacuum{}
\def\draftmarginnote#1{\marginpar{\raggedright\scriptsize\tt#1}}
\def\draft{\oddsidemargin 0.0truein
\def\@oddfoot{\sl ES, preliminary notes \hfil
\rm\thepage\hfil\sl\today\quad\militarytime}
\let\@evenfoot\@oddfoot \overfullrule 3pt
\let\label=\draftlabel
\let\marginnote=\draftmarginnote
\def\@eqnnum{(\theequation)\rlap{\kern\marginparsep\tt\@eqnlabel}%
\global\let\@eqnlabel\@vacuum} }
\catcode`@=12
\newcommand{\be}{\begin{equation}}
\newcommand{\ee}{\end{equation}}

\def\({\left(}
\def\){\right)}
\def\<{\langle}
\def\>{\rangle}

\def\be{\begin{equation}}
\def\ee{\end{equation}}
\def\bry{\begin{array}}
\def\ery{\end{array}}
\def\bes{\begin{subequations}}
\def\ees{\end{subequations}}
\def\bit{\begin{itemize}}
\def\eit{\end{itemize}}
\def\ben{\begin{enumerate}}
\def\een{\end{enumerate}}

\newcommand{\MET}{E\llap{/\kern1.5pt}_T}
\definecolor{grey}{rgb}{0.6,0.6,0.6}
\definecolor{fuchsia}{rgb}{1,0,1}

\makeatother
\begin{document}

\preprint{CPHT-RR085.0913}
\preprint{DFPD-2013/TH/17}
\preprint{LPN13-058}

\title{Collider signatures of low scale supersymmetry breaking: A Snowmass 2013 White Paper}
\author{Emilian Dudas}
\email{emilian.dudas@cpht.polytechnique.fr}
\address{Centre de Physique Th\'eorique, \'Ecole Polytechnique, CNRS, Palaiseau, France}
\author{Christoffer Petersson}
\email{christoffer.petersson@ulb.ac.be}
\address{Physique Th\'eorique et Math\'ematique, Universit\'e Libre de Bruxelles, C.P. 231, 1050 Bruxelles, Belgium}
\address{International Solvay Institutes, Brussels, Belgium}
\address{Department of Fundamental Physics, Chalmers University of Technology, 412 96 G\"oteborg, Sweden}
\author{Riccardo Torre}
\email{riccardo.torre@pd.infn.it}
\address{Dipartimento di Fisica e Astronomia, Universit\`a di Padova, and \\ INFN Sezione di Padova, Via Marzolo 8, I-35131 Padova, Italy}
\address{SISSA, Via Bonomea 265, I-34136 Trieste, Italy}
\begin{abstract}
We consider the possibility that supersymmetry is broken at a low scale, within one order of magnitude above the TeV scale. In such a case, the degrees of freedom associated with the spontaneous breaking of supersymmetry, the goldstino fermion and its scalar superpartner (the sgoldstino), can have significant interactions with the Standard Model particles and their superpartners. We discuss some characteristic processes, involving the goldstino and the sgoldstino, and the collider signatures they give rise to. These signatures involve scalar resonances in the di-photon, di-jet, di-boson and di-tau channels, the possible relation between these resonances and deviations in the Higgs couplings as well as exotic Higgs decays in the monophoton\,+$\MET$ and the four photon channels.
\end{abstract}
%
%

\maketitle

\section{Introduction}
\vspace{-2mm}

If supersymmetry (SUSY) is a feature of Nature it must be in a broken phase at low energies. A model-independent consequence of the spontaneous breaking of (global) SUSY is the existence of a Goldstone Weyl fermion, the goldstino. The existence of gravity implies that SUSY is a local symmetry, that the spin $1/2$ goldstino is eaten by the spin $3/2$ gravitino, becoming its longitudinal component, and that the gravitino acquires a mass $m_{3/2}= f/(\sqrt{3}M_{\mathrm{P}})$, where $\sqrt{f}$ is the SUSY breaking scale and $M_{\mathrm{P}}=2.4\cdot 10^{18}$\,GeV. 

Treating $\sqrt{f}$ (or, equivalently, $m_{3/2}$) as a free parameter one can categorize different  scenarios of SUSY breaking and mediation. Gravity and anomaly mediation are relevant for high scale SUSY breaking, with $\sqrt{f}$ of the order of $10^{11}$\,GeV, while gauge mediation is relevant for lower values of $\sqrt{f}$, down to around 50 TeV (below which one typically encounters tachyonic scalars in the messenger sector). However, the lower experimental bound on $\sqrt{f}$ is at or below 1 TeV, which leaves the possibility of having $\sqrt{f}$ within an order of magnitude above the TeV scale, which is the case we discuss in this note. When $\sqrt{f}$ is of this order, the gravitino is approximately massless and, due to the supersymmetric equivalence theorem \cite{1977PhLB...70..461F,1988PhLB..215..313C}, the gravitino can be replaced by its goldstino components. For different aspects and discussions on models with a low SUSY breaking scale see, \mbox{for example, Refs.~\cite{Brignole:1997hb,Djouadi:1997bx,1997hep.ph...11516B,Brignole:1998eg,Perazzi:2000ku,Gherghetta:2000br,Perazzi:2000dk,Gorbunov:2000ii,Gorbunov:2002co,Brignole:2003hb,Antoniadis:2010hs,Azatov:2011dn,Dudas:2011wk,Gherghetta:2011tn,Antoniadis:2011ve,Bertolini:2011wj,Petersson:2011in,Petersson:2012tl,deAquino:2012vu,Bellazzini:2012ul,Antoniadis:2012ui,2013PhRvD..87a3008P,Riva:2012ti,Dudas:2012ti,Farakos:2013wo,Christensen:2013ve}.}

We consider the case where the goldstino  $\widetilde{G}$ resides in a gauge singlet chiral superfield
\begin{equation}
\label{X}
X=x+\sqrt{2}\theta \widetilde{G}+\theta^2 F_X\,,
\vspace{-1mm}
\end{equation}
where the auxiliary component $F_X$ acquires a non-vanishing vacuum expectation value ({\scshape vev}), $\langle F_X \rangle=f$, that breaks SUSY.\footnote{In general the goldstino is a linear combination of all the neutral fermions whose associated auxiliary $F$ or $D$ component acquires a non-vanishing {\scshape vev}. Here we assume that $| \langle F_X \rangle |^2$ provides the dominant contribution to the vacuum energy and that the goldstino, to a good approximation, is aligned with the fermion component of $X$.} For linearly realized (but spontaneously broken) SUSY, the goldstino has a complex scalar superpartner, the sgoldstino $x$.\footnote{Since the sgoldstino is complex it gives rise to two real scalars, one $CP$-even and one $CP$-odd. In terms of R-parity, the goldstino is odd while the sgoldstino is even.} In contrast to the goldstino, the sgoldstino is not protected by the Goldstone shift symmetry and it generically acquires a mass upon integrating out some heavy states in the SUSY breaking sector. The precise value of its mass depends on the details of that sector, such as the symmetries (e.g.~R-symmetry) and the loop level at which its mass is generated. If the sgoldstino mass is much larger than the energy scale under consideration, it can be integrated out and as a consequence, SUSY is non-linearly realized in the resulting low energy effective theory. In the case where the SUSY breaking sector is strongly coupled and no energy scale exists at which SUSY is linearly realized, the elementary scalar component in Eq.~\eqref{X} is replaced by a goldstino bilinear, $x\to \widetilde{G}\widetilde{G}/(2F_X)$, such that the corresponding non-linear superfield $X_{NL}$ satisfies the constraint $X_{NL}^2=0$ \cite{Komargodski:2009cq}.

Being a Goldstone mode, the goldstino couples derivatively to the supercurrent, or, upon using the equations of motion, non-derivatively to the divergence of the supercurrent. A simple way to incorporate the interactions of the goldstino, and its sgoldstino superpartner, with the Standard Model (SM) fields and their superpartners is to promote all soft terms to SUSY operators containing the goldstino superfield of Eq.~\eqref{X}, 
\vspace{-2.5mm}
\begin{eqnarray}
 m_{\widetilde{\phi}}^2 \widetilde{\phi}^\dagger \widetilde{\phi} & \to & \int d^4\theta \frac{m_{\widetilde{\phi}}^2}{f^2} X^\dagger X \Phi^\dagger \Phi \,,\label{scalar}\\
m_\lambda \lambda^\alpha\lambda_\alpha & \to &  \int d^2\theta \frac{m_\lambda}{f} X W^\alpha W_\alpha \,,
\label{gaugino}\vspace{-2.5mm}
\end{eqnarray}
where $\Phi$/$W^\alpha$ are generic matter/gauge superfields and $m_{\widetilde{\phi}}^2$/$m_\lambda$ are the corresponding scalar/gaugino soft masses. In addition to the soft mass terms, which are recovered from the SUSY operators by taking the auxiliary components of the $X$'s and inserting their {\scshape vev}'s, the SUSY operators in Eqs.~\eqref{scalar} and \eqref{gaugino} also give rise to goldstino and sgoldstino interactions with the component fields of the matter/gauge supermultiplets. The coefficients of these interactions are proportional to the ratio of the corresponding soft parameter over the SUSY breaking scale. Hence, in the scenario where the $\sqrt{f}$ is not far above the scale of the soft parameters, the goldstino and sgoldstino interactions can be significant.

Note that the auxiliary component $F_X$ of the goldstino superfield of Eq.~\eqref{X} is treated dynamically and upon integrating it out, it gives rise to a tree level $F$-term scalar potential. As can be seen from Eq.~\eqref{scalar} in the case where $\Phi$ is a Higgs superfield, $F_X$ couples to the Higgs scalars, which gives rise to new quartic Higgs couplings in the $F$-term scalar potential, beyond those proportional to gauge coupling constants in the usual $D$-term scalar potential. These new interactions can help to raise the mass of the lightest neutral CP-even Higgs scalar to 125 GeV already at the tree level \cite{Antoniadis:2010hs,Petersson:2011in}.

\vspace{-3mm}
\section{Bounds on the SUSY breaking scale}
\vspace{-2mm}
Model-independent bounds on the SUSY breaking scale/gravitino mass can be derived by means of an effective approach in which all the SM superpartners are taken to be heavy and integrated out from the spectrum. In this case, effective higher dimensional operators give rise to the processes $e^{+}e^{-}\to \widetilde{G}\widetilde{G}\gamma$ at $e^{+}e^{-}$ colliders and $pp\to \widetilde{G}\widetilde{G}\gamma$ and $pp\to \widetilde{G}\widetilde{G}j$ at hadron colliders \cite{1997hep.ph...11516B,Brignole:1998eg}. These effective operators are suppressed by powers of the SUSY breaking scale and therefore these processes can be used to set a lower bound on $\sqrt{f}$. 

Several experimental searches have been performed for these signatures, referred to respectively as monophoton plus missing energy ($\gamma+\MET$) and monojet plus missing energy ($j+\MET$), at LEP, Tevatron and LHC. The most updated bound from the LEP experiments comes from the L3 Collaboration \cite{LCollaboration:2004hl}, where a conservative bound of \mbox{$\sqrt{f}>238$ GeV} was obtained. The CDF Collaboration at the Tevatron has looked both for the $\gamma+\MET$ \cite{2002hep.ex....5057C} and the $j+\MET$ \cite{CDFCollaboration:2000bc} signatures and the resulting bounds are comparable to the LEP one: $\sqrt{f}>221 $ GeV and $\sqrt{f}>214$ GeV, respectively. 

The LHC experiments have also performed searches for both $\gamma+\MET$ and $j+\MET$, but with focus mainly on large extra dimension and dark matter models.\footnote{See Ref.~\cite{Dudas:2012ti} for a discussion concerning LHC bounds on four-fermion contact interactions and how they can be translated into bounds on higher-dimensional operators, present in low scale SUSY breaking models, with coefficients involving the SUSY breaking scale.} So far only one analysis, done by the ATLAS Collaboration, has presented their results in terms of bounds on $\sqrt{f}$ \cite{ATLAS-CONF-2012-147}. In this analysis the superpartners are not integrated out and different relations between the squark and gluino masses are considered. The most conservative bound that can be extracted from Ref.~\cite{ATLAS-CONF-2012-147} is $\sqrt{f}>650$ GeV for heavy squarks and gluinos but bounds as strong as $\sqrt{f}\gtrsim 1.1$ TeV are obtained for particular relations between the squark and gluino masses.

\vspace{-3mm}
\section{Signatures of low scale SUSY breaking models}
\vspace{-2mm}
In this section we briefly review some of the characteristic signatures of models with a low SUSY breaking scale, arising from processes involving the goldstino and the sgoldstino. 
\vspace{-4mm}
\subsection{Sgoldstino production and decay}
\vspace{-3mm}

By taking the scalar sgoldstino component from the goldstino superfield $X$ and the gauge kinetic term $F_{\mu\nu}F^{\mu\nu}$ from $W^\alpha W_\alpha$ in \eqref{gaugino} we see that the sgoldstino couples to (the transverse components of) the SM gauge bosons. The strengths of these interactions are given by the corresponding gaugino soft mass over $f$. For example, the sgoldstino coupling to gluons is proportional to the gluino mass and the coupling to photons is given by a linear combination (with weak mixing angles) of the bino and wino masses. At a hadron collider the sgoldstino is resonantly produced via gluon-gluon fusion. The production cross-section can be found in Ref.~\cite{Bellazzini:2012ul}. Since the sgoldstino can couple strongly to gauge bosons, relevant searches for sgoldstinos involve direct searches for a scalar resonance into the di-photon ($x\to \gamma\gamma$), di-jet ($x\to gg$) or di-boson ($x\to ZZ$, $x\to WW$) final states. Due to the very high background in the di-jet final state at the LHC, we expect the di-boson searches to be more efficient in the low mass region ($\lesssim 1$ TeV). Searches in the aforementioned final states are often performed by the experimental collaborations having in mind different kind of resonances, such as graviton excitations in models with extra dimensions or sequential SM $Z'$ and $W'$ (see, e.g., Refs.~\cite{ATLASCollaboration:2012ka,CMSCollaboration:2012by}). However, it would be very useful if the experimental efficiencies and the kinematic acceptances were given also for scalar particles in these searches. Moreover, while in the $WW$ and $ZZ$ final states the whole region starting from $m_{h}$ and up to a few TeV seems to be covered by the experimental analyses (Higgs searches plus Exotics searches), there appears to be a gap in the searches in the di-photon channel for resonances with a mass between 150 GeV (the upper end of the reach in the Higgs search in the di-photon channel) and 500 GeV (the lower end in the search for Kaluza-Klein gravitons \cite{ATLASCollaboration:2012ka}).

\vspace{-4mm}  
\subsection{Implications for Higgs Couplings}
\vspace{-3mm}

As was discussed above, the sgoldstino scalar can couple strongly to the SM gauge bosons. Moreover, the \mbox{CP-even} sgoldstino will in general mix with the scalar state corresponding to the SM-like Higgs at 125 GeV. Therefore, through mixing with sgoldstino, the SM-like Higgs can have modified couplings to gauge bosons. In Ref.~\cite{Bellazzini:2012ul} it was shown that, through a tree-level sgoldstino mixing correction, it is possible to enhance the Higgs coupling to photons without significantly modifying any of the other Higgs couplings. If the Higgs coupling to photons would turn out to be enhanced with respect to the SM expectation, in the context of this scenario, it would suggest that the sgoldstino couples strongly to photons and that a scalar resonance in the di-photon channel below the TeV scale would be well-motivated to search for. Note that the sgoldstino mixing is uniquely determined by the soft parameters, see Refs.~\cite{Dudas:2012ti,2013PhRvD..87a3008P} for the explicit mixing corrections to all the couplings of the SM-like Higgs scalar.

In Ref.~\cite{2013PhRvD..87a3008P} it was studied the possibility of using sgoldstino mixing to disentangle the usual relations between the different Higgs couplings to SM fermions.\footnote{In addition to the sgoldstino mixing corrections, modifications of the Higgs couplings to fermions arise from, for instance, ``wrong Yukawa couplings", e.g.~$X^\dagger H_d^\dagger Q U^c$ in the Kahler potential, which can be generated and significant in models with a \mbox{low SUSY breaking scale \cite{Dudas:2012ti}.}\vspace{1mm}} In the minimal SUSY SM (MSSM), the Higgs couplings to the down-type quarks and the leptons, normalized with respect to their corresponding SM value, coincide at the tree-level and this degeneracy is typically only slightly broken at the quantum level. The sgoldstino couples to the SM fermions via superpotential operators such as $(A_d/f)XQH_d D^c$ and $(A_e/f)XLH_dE^c$, which also give rise to the tri-linear scalar soft $A$-terms. Since the sgoldstino mixing corrections to the Yukawa couplings depend on the separate soft $A$-term for the corresponding fermion, this allows for the freedom to break the usual relations between the different Higgs couplings to fermions. If, for example, the Higgs coupling to tau leptons is shown to deviate from the SM expectation, but not the Higgs coupling to bottom quarks, and if a sgoldstino mixing correction would be responsible for this deviation, it would imply that the sgoldstino itself couples strongly to taus. In this case, a search for a scalar resonance in the di-tau channel would be well-motivated. The most relevant searches in the di-tau channel are given by Refs.~\cite{CMSCollaboration:2012ek,ATLASCollaboration:2012un} which, however, focus on $Z'$ resonances and do not not provide the efficiency times acceptance for a scalar particle. As it was already pointed out for the di-boson case, it would be very useful to have these quantities for a scalar particle from the experimental collaborations in order to be able to interpret the analyses in terms of a low scale SUSY breaking model with a sgoldstino scalar in the low energy spectrum.

\vspace{-4mm}
\subsection{Exotic Higgs decays}
\vspace{-3mm}
In this section we discuss exotic Higgs decays, involving the goldstino and the sgoldstino, for  which we would like to encourage experimental searches. \vspace{2mm}

\noindent $\bullet$ $h\to\gamma+\MET$
\vspace{1mm}

 \noindent This Higgs decay into a monophoton plus missing transverse energy arises from the process \mbox{$h\to \widetilde{\chi}^0_1 \widetilde{G}\to \gamma \widetilde{G}\widetilde{G}$} which is kinematically allowed when the lightest neutralino is lighter than the Higgs. The coupling $h\widetilde{\chi}^0_1\widetilde{G}$ arises from, for instance, Eq.~\eqref{scalar} (with $\Phi$ being a Higgs superfield), by taking the auxiliary field (and inserting its {\scshape vev}) from one of the $X$'s and the goldstino component from the other $X$, while taking a Higgs scalar and a higgsino component, respectively, from the two Higgs superfields. The $\widetilde{\chi}^0_1\gamma\widetilde{G}$ coupling arises from Eq.~\eqref{gaugino} by taking the goldstino component from $X$, while taking a gaugino and a gauge field strength component from $W^\alpha W_\alpha$.\footnote{If kinematically allowed, also the (subleading) process \mbox{$h\to \widetilde{\chi}^0 \widetilde{G}\to Z \widetilde{G}\widetilde{G}$} occurs \cite{Petersson:2012tl}.}  Since the kinematic distribution of the transverse momentum of the photon has an endpoint at half the Higgs mass, the relevant phase space regime is \mbox{$p_T^\gamma\leqslant m_h/2\approx 63$ GeV}. This process, as well as the relevant backgrounds, were studied in Ref.~\cite{Petersson:2012tl}.\vspace{2mm}

\noindent $\bullet$  $h\to 4\gamma/2\gamma2j/4j$\vspace{1mm}

 \noindent  These Higgs decays arise from the processes \mbox{$h\to xx\to(\gamma\gamma)(\gamma\gamma)/(\gamma\gamma)(jj)/(jj)(jj)$}, which are present when the sgoldstino mass is below half the Higgs mass.\footnote{The 4 jet final state is expected to be swamped by the QCD background at a hadron collider. However, if one or both of the sgoldstino scalars decay to $b$-quarks, then $b$-tagging and suitable angular variables can help increasing the sensitivity in the $2b2j$ and $4b$ \mbox{final states \cite{Franceschini:2012vl}.}}  In the particular case where the sgoldstinos are very light (100-400 MeV) the analysis of Ref.~\cite{ATLAS-CONF-2012-079} is sensitive since it searches for Higgs decays into two pseudo-scalars, each of which decays to two photons. Due to the strong boost, each of the two pairs of photons are so collimated that most of them are misidentified with single photons and hence this decay contributes to the usual di-photon channel. However, in the case where the sgoldstino is heavier than that it should be possible to isolate and reconstruct all the 4 photons. The decays of the Higgs into $4\gamma/2\gamma2j$ via two pseudo-scalars in the context of a non-SUSY model with a pseudo-scalar and heavy vector-like fermions were considered in Refs.~\cite{2001PhRvD..63g5003D,Chang:2006bw}. \vspace{2mm}

\noindent $\bullet$  $h\to \MET$\vspace{1mm}

\noindent This invisible Higgs decay can arise from the process \mbox{$h\to \widetilde{G}\widetilde{G}$}, see for instance Ref.~\cite{Antoniadis:2010hs}. 
 
\vspace{-3mm}
\section{Conclusions}
\vspace{-2mm}
Bounds on the scale of SUSY breaking are currently at or below 1 TeV. With no theoretical bias, and motivated by the negative results of the SUSY searches in the first run of the LHC, we point out that models where SUSY is broken around the TeV scale can have non-standard and distinctive phenomenology. New processes and modifications to the SM ones arise in this framework, depending only on the soft parameters and the SUSY breaking scale. 

The characteristic processes we have discussed involve the degrees of freedom associated with the spontaneous breaking of SUSY, i.e.~the goldstino and the sgoldstino. In contrast to other neutral fermions and scalars present in other extensions of the SM, the couplings of the goldstino/sgoldstino to the SUSY SM are proportional to the soft parameters and masses of the SM superpartners. Hence, these couplings contain information about the superpartner spectrum that can be used to relate seemingly disconnected experimental analyses, such as searches for superpartners, new scalar resonances, exotic Higgs decays and modified Higgs couplings. 

Some of the distinctive signatures described in this note have analogous signatures in other models, for which analyses are already available. In this case it would be very useful if the experimental collaborations could provide the efficiencies and acceptances for all the different spin/Lorentz structures that could give rise to a given signature. 

In conclusion, we would like to encourage experimental searches for models with a low SUSY breaking scale. Moreover we think that a systematic experimental program for this framework should be established. \\

\vspace{-6.5mm}
\section*{Acknowledgements}
\vspace{-2mm}
The  work of E.D. was supported in part by the
European ERC Advanced Grant 226371 MassTeV, the French ANR TAPDMS 
ANR-09-JCJC-0146 and
the contract PITN-GA-2009-237920 UNILHC. 
The work of C.P. was supported in part by IISN-Belgium (conventions 4.4511.06, 4.4505.86 and 4.4514.08), by the ``Communaut\'e Fran\c{c}aise de Belgique" through the ARC program and by a ``Mandat d'Impulsion Scientifique" of the F.R.S.-FNRS. 
The work of R.~T. was supported by the ERC Advanced Grant no. 267985 {\it DaMeSyFla}. R.T. also acknowledges the Grant Agreement number PITN-GA-2010-264564 {\it LHCPhenoNet} of the Research Executive Agency (REA) of the European Union.

\bibliographystyle{mine}
\vspace{-3mm}
\bibliography{bibliography}

\end{document}